\documentclass[a4paper]{mem}
\usepackage{natbib}
\usepackage{graphicx}
\usepackage{amssymb}
\usepackage[a4paper]{hyperref}
\idline{73}{23}

\begin{document}
   \title{A Search for Low Mass Stars and Brown Dwarfs in the Upper Scorpius OB Association%
}

   \author{C.L. Slesnick, J.M. Carpenter \and L.A. Hillenbrand 
}

   \offprints{C.L. Slesnick}
\mail{cls@astro.caltech.edu}

   \institute{Dept. of Astronomy, MS105-24, California Institute of Technology,Pasadena, CA 91125}

  \abstract{
We are conducting a $U,B,R,I$ monitoring program to identify pre-main sequence 
stars and brown
dwarfs in a 12.5$^o\times12.2^o$ region of the Upper Scorpius OB association  (5--10 Myr).
  We
will use these data in combination with a follow-up spectroscopy survey to derive the 
low mass IMF in Upper Sco and to explore the prevalence of 5--10 Myr circumstellar disks. 
We will also analyze the
spatial distribution of association members as a function of stellar mass,
from which we will be able to 
place constraints on brown dwarf formation scenarios.  We expect to identify 700--1800
previously unknown pre-main sequence objects in Upper Sco with M $<$ 0.6 M$_\odot$.
This dataset will constitute the largest 
sample known to date of stars/brown dwarfs at 5--10 Myr.

   \keywords{Brown dwarfs -- Variability -- Clusters -- Circumstellar disks}
}

  \authorrunning{Slesnick et al.}
   \titlerunning{USco Variability Survey}
   \maketitle
%

%

\section{Introduction}

   \begin{figure*}
   \centering
   \scalebox{0.55}{\includegraphics{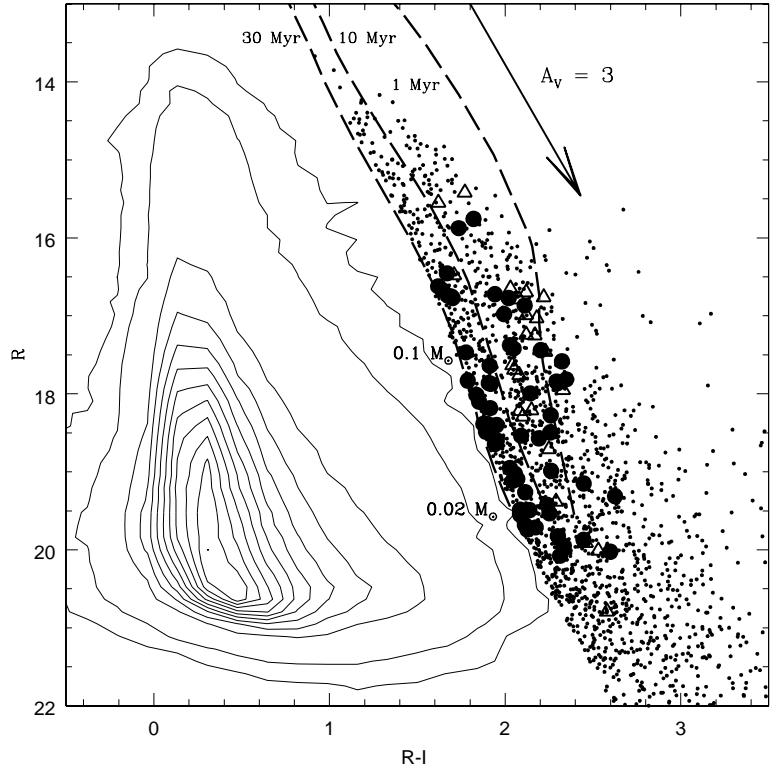}}
   \caption{$R,(R-I)$ color-magnitude diagram for the Upper Sco survey data.  Contours indicate the density of sources detected in
both $R$-- and
$I$--bands ($\sim$1 million with good photometry in both bands).  Data are represented at the
90\%,80\%,70\%,60\%,50\%,40\%,30\%,20\%,10\%,1\% and 0.5\% levels.
Objects which appear younger than 30 Myr based on their location on the color-magnitude diagram are shown as discrete points.
Stars identified as variables are drawn as large circles.  Spectroscopically observed Upper Sco members from
Ardila et al. (2000) are shown as open triangles.  We expect low mass cluster members to exhibit only
moderate extinction ($A_V <$ 1; Mart\'{i}n et al. 2004).}
              \label{Figcmd}%
    \end{figure*}

Understanding the formation and evolution of young brown dwarfs and low mass stars
requires large samples of objects  
spanning a wide range in stellar mass and age.
One difficulty faced by studies in these areas is that while 
hundreds of nearby ($<$ 140~pc) stars and brown dwarfs have been identified at ages of 
$<$ 1-3~Myr
and 
$\sim$100~Myr, less than a few tens 
have been identified at ``intermediate'' ages ($\sim$3-30~Myr; eg., TW Hydrae and $\beta$ Pic moving groups). 
Determining the properties of low mass objects 
and, when present, their 
associated circumstellar disks at these intermediate ages
is crucial to our understanding of many aspects of star and planet formation. 
In particular, recent work by Moraux \& Clarke (2004) 
suggests that the 
spatial
distributions of low mass members in 
intermediate-age clusters (age $ \sim$one crossing time)
can be used as a discriminator between 
gravitational collapse and ejection
formation scenarios for low mass objects.
In addition, recent 
near-infrared surveys of young ($<$3 Myr) star-forming clusters 
(eg., IC 348 \& Taurus; Liu, Najita \& Tokunaga 2003) find
evidence of high disk frequencies ($\approx$80\%) for substellar objects. 
Exploring more evolved 5--30 Myr counterparts to these young disks 
is essential given that theoretical and 
observational considerations suggest planets are in their 
final assemblage stages during this evolutionary period. 
   \begin{figure*}
   \centering
   \scalebox{0.6}{\includegraphics{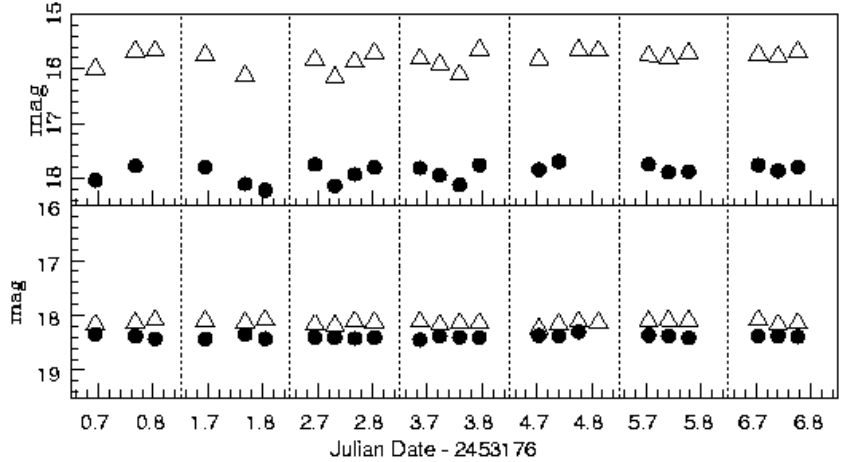}}
   \caption{
Upper panels show the light curve of a known low 
mass pre-main sequence star in Upper Sco 
(UScoCTIO 100; Ardila et al. 2000) which we found to be variable in our Quest-2 data.  
Lower panels are the light curve for a nearby ($<$20 arcsec) object, shown to demonstrate
the quality of the photometry.  In all panels open triangles are $I$-band data and circles are $R$-band data. 
Dotted lines represent gaps between nights in the time series.
}
              \label{Figcmd}%
    \end{figure*}

The lack of intermediate 
age objects in current samples is certainly an observational bias.
Most nearby known intermediate-aged populations are comprised of
moving groups or OB associations which cover large areas on the sky,
 and whose low mass pre-main sequence populations are difficult to distinguish
from field stars.
Upper Sco is the closest (140 pc; de Zeeuw
et al. 1999) young OB association to the Sun with  
114 known high mass Hipparcos stars.
At $\sim$5 Myr (Preibisch et al. 2002), this cluster is at a critical age 
between very young ($<$3 Myr) star forming regions and older ($>$100 Myr) Pleiades-type open clusters.
Spectroscopic surveys covering large areas but still relatively small compared to 
the overall size of the OB association by
Preibisch et al. (2001, 2002) and Mart\'{i}n et al. (2004) identified $\sim$200 low mass pre-main sequence stars and brown dwarfs (see also Meyer et al. 1993, Walter et al. 1994, Mart\'{i}n 1998, Preibisch et al. 1998 and Ardila et al. 2000).
Assuming that the Upper Sco initial mass function is representative of other
star forming regions, the association should contain thousands of low mass
objects which are yet to be discovered against the back drop of field stars.

\section{Current Observations and Pre-Main Sequence Candidate Selection}
To rectify this situation we are conducting a $U,B,R,I$ monitoring program 
to identify the low mass pre-main sequence objects in Upper Sco. 
Our survey data are taken with the Quest-2 camera (Rabinowitz et al. 2003) on the Palomar 48-inch telescope 
in driftscan format over
$\approx$12.5$^o$$\times$12.2$^o$ (encompassing $\sim$65\% of the known Hipparcos members
of Upper Sco), with part of the region
 scanned 3--4 times per night on 7 consecutive nights.
The large spatial coverage of our survey
includes both the known population of 5 Myr objects and stars 
that may have formed in the central regions of Upper Sco within
the past $\approx$5--30~Myr and subsequently drifted outside.  

Our final dataset contains $\sim$8 million sources
from which we have selected $\sim$0.1\% of the data as candidate pre-main sequence objects
based on $R,(R-I)$ magnitudes and colors indicating 
they could be young ($<$ 10 Myr) Upper Sco members (see Fig. 1).
When possible,
further refinement of our candidate list will be made via analysis of 
2MASS 
near-infrared magnitudes/colors.
In addition, we have photometric monitoring of
the central 12.5$^o$$\times$4.6$^o$ strip which allows us to identify variable stars and 
study activity
of pre-main sequence candidates.
This activity
occurs through a number of processes, the two most prominent 
being related to residual accretion of material 
from circumstellar disks onto the photospheres (ages $\lesssim$3 Myr) 
and chromospheric activity
(ages $\lesssim$ 120 Myr). 
Activity signatures are observable through several means, most notably photometric variability,
UV excess and X-ray emission. 
Of these choices, photometric variability is the most efficient means of identifying large numbers of 
low mass (faint) pre-main-sequence stars 
over a broad range of ages. 
From the photometric monitoring data we have selected $\sim$500 objects which exhibit 
variability on timescales of hours/days (see Fig. 2).

\section{Future Observations and Analysis}
We are in the process of obtaining intermediate resolution follow-up optical spectroscopy
to confirm youth and spectral type (leading to mass) of candidate pre-main sequence stars and brown dwarfs selected from photometric variability and/or location on an optical color-magnitude diagram.
Observing color-magnitude diagram-selected and variability-selected objects will enable us to measure the completeness 
of the two techniques.
We will measure the shape and equivalent width of H$\alpha$ in our observed spectra 
which will allow
us to assess the origin of an object's variability.  Broad, asymmetric profiles 
arise from 
active accretion from an optically thick circumstellar disk whereas narrow, symmetric profiles are indicative of 
chromospheric activity.  
This technique has been applied successfully to a small subset of previously known low mass 
Upper Sco members (Muzerolle et al. 2003; Jayawardhana et al. 2003).  
Our larger sample will allow us to derive more robust statistics for  
circumstellar disks within the cluster as a function of mass and age.

We will also analyze the mass, age and spatial distributions of spectroscopically
confirmed Upper Sco members.
Preibisch et al. (2002) derived an IMF for Upper Sco from $\sim$166 stars 
(20 M$_\odot$ $>$ M $>$ 0.1 M$_\odot$) which showed some evidence for an excess of 
low mass stars in comparison to the higher mass cluster members.  
We will use our extensive catalog of $\sim$700-1800 cluster members to extend 
this work into the substellar regime (0.6 M$_\odot$ $>$ M $>$ 0.02 M$_\odot$).  
The spatial distribution of very low mass cluster members within the 
$\sim$150 deg$^2$ survey area will be assessed
and used 
to place constraints on brown dwarf formation scenarios.
Mass segregation in a cluster at 5 Myr could not be caused by dynamical 
relaxation but instead might be attributed to the lowest mass members 
having been ejected with systematically larger spatial velocities (as compared to the higher
mass stars) from collapsing systems.  Conversely,
if we find no evidence for differences in the spatial distributions of low
and high mass cluster members, it will provide a very strong
argument in favor of both low and high mass objects forming 
via a scenario similar to the standard star formation model.

\end{document}